\documentclass[12pt]{article}
\input epsf
\usepackage{amssymb}
\usepackage{psfrag}
\usepackage{graphicx}

\renewcommand{\a}{\alpha}
\renewcommand{\b}{\beta}

\newcommand{\rmd}{{\rm d}}
\newcommand{\m}{\mu}
\newcommand{\n}{\nu}

\def\be{\begin{equation}}
\def\ee{\end{equation}}
\def\bea{\begin{eqnarray}}
\def\eea{\end{eqnarray}}
\def\ba{\begin{array}}
\def\ea{\end{array}}
\def\bi{\begin{itemize}}
\def\ei{\end{itemize}}
\def\half{{\textstyle{1\over2}}}

\catcode`\@=11
\def\@citex[#1]#2{%
\if@filesw \immediate \write \@auxout {\string \citation {#2}}\fi
\@tempcntb\m@ne \let\@h@ld\relax \def\@citea{}%
\@cite{%
  \@for \@citeb:=#2\do {%
    \@ifundefined {b@\@citeb}%
      {\@h@ld\@citea\@tempcntb\m@ne{\bf ?}%
      \@warning {Citation `\@citeb ' on page \thepage \space undefined}}%
      {\@tempcnta\@tempcntb \advance\@tempcnta\@ne%
      \@tempcntb\number\csname b@\@citeb \endcsname \relax%
      \ifnum\@tempcnta=\@tempcntb 
        \ifx\@h@ld\relax%
          \edef \@h@ld{\@citea\csname b@\@citeb\endcsname}%
        \else%
          \edef\@h@ld{\ifmmode{-}\else--\fi\csname b@\@citeb\endcsname}%
        \fi%
      \else
        \@h@ld\@citea\csname b@\@citeb \endcsname%
        \let\@h@ld\relax%
      \fi}%
    \def\@citea{,\penalty\@highpenalty\,}%
  }\@h@ld
}{#1}}

\def\@citeb#1#2{{[#1]\if@tempswa , #2\fi}}
\def\@citeu#1#2{{$^{#1}$\if@tempswa , #2\fi }}
\def\@citep#1#2{{#1\if@tempswa , #2\fi}}

\def\bcites{         
        \catcode`\@=11
        \let\@cite=\@citeb
        \catcode`\@=12
}

\def\upcites{         
        \catcode`\@=11
        \let\@cite=\@citeu
        \catcode`\@=12
}

\def\plaincites{      
        \catcode`\@=11
        \let\@cite=\@citep
        \catcode`\@=12
}

\newcount\hour
\newcount\minute
\newtoks\amorpm
\hour=\time\divide\hour by 60
\minute=\time{\multiply\hour by 60 \global\advance\minute by-\hour}
\edef\standardtime{{\ifnum\hour<12 \global\amorpm={am}%
        \else\global\amorpm={pm}\advance\hour by-12 \fi
        \ifnum\hour=0 \hour=12 \fi
        \number\hour:\ifnum\minute<10 0\fi\number\minute\the\amorpm}}
\edef\militarytime{\number\hour:\ifnum\minute<10 0\fi\number\minute}

\def\draftlabel#1{{\@bsphack\if@filesw {\let\thepage\relax
   \xdef\@gtempa{\write\@auxout{\string
      \newlabel{#1}{{\@currentlabel}{\thepage}}}}}\@gtempa
   \if@nobreak \ifvmode\nobreak\fi\fi\fi\@esphack}
        \gdef\@eqnlabel{#1}}
\def\@eqnlabel{}
\def\@vacuum{}
\def\marginnote#1{}
\def\draftmarginnote#1{\marginpar{\raggedright\scriptsize\tt#1}}
\def\draft{
        \pagestyle{plain}
        \overfullrule=2pt
        \oddsidemargin -.5truein
        \def\@oddhead{\sl \phantom{\today\quad\militarytime} \hfil
        \smash{\Large\sl DRAFT} \hfil \today\quad\militarytime}
        \let\@evenhead\@oddhead
        \let\label=\draftlabel
        \let\marginnote=\draftmarginnote
        \def\ps@empty{\let\@mkboth\@gobbletwo
        \def\@oddfoot{\hfil \smash{\Large\sl DRAFT} \hfil}
        \let\@evenfoot\@oddhead}
        \def\@eqnnum{(\theequation)\rlap{\kern\marginparsep\tt\@eqnlabel}%
        \global\let\@eqnlabel\@vacuum}  }

\begin{document}

\hfill CERN-TH/2001-310

\hfill UTHET-01-1101

\hfill {\tt hep-th/0111085} 
\vspace{-0.2cm}

\begin{center}
\Large
{ \bf dS/CFT correspondence on a brane}
\normalsize

\vspace{0.8cm}
{\bf
A. C. Petkou}\footnote{tassos.petkou@cern.ch} \\ 
CERN Theory Division, \\
CH 1211 Geneva 23, \\
Switzerland
\vspace{.5cm}
 
{\bf G. Siopsis}\footnote{gsiopsis@utk.edu}\\ Department of Physics
and Astronomy, \\
The University of Tennessee, Knoxville \\
TN 37996 - 1200, USA.
 \end{center}

\vspace{0.8cm}
\large
\centerline{\bf Abstract}
\normalsize
\vspace{.5cm}

We study branes moving in an AdS Schwarzschild black hole
background. When the brane tension exceeds a critical value, the induced
metric on the brane is of FRW type and asymptotically de Sitter. We
discuss the relevance of such configurations to dS/CFT
correspondence. When the black hole mass reaches a critical value that
depends on the brane tension, the brane interpolates
in the infinite past and future between a dS space and a finite space
of zero Hubble constant. This corresponds to a
cosmological evolution  without 
a Big Bang or a Big Crunch. Moreover,
the central charge of the CFT dual to the dS brane enters the Cardy-Verlinde
formula that gives the entropy of the thermal CFT dual to the bulk AdS
black hole. 

\newpage

\section{Introduction}

The AdS$_{D+1}$/CFT$_D$ correspondence \cite{maldacena} is nowadays
largely accepted as  an 
established duality, holographical in nature,  between a 
$D+1$-dimensional gravitational theory in the bulk and a
$D$-dimensional CFT in the boundary. A
generalisation of holographic ideas leads naturally to conjecture the existence
of a dS$_{D+1}$/CFT$_D$  correspondence. In this case, holography
takes place on the asymptotic boundaries
along the time 
direction and the dual CFT (CFTs) is (are)
Euclidean. Despite some serious puzzles, 
most notably 
the unclear nature of a string realization of the dS/CFT correspondence, such a
proposal has been recently put forward in a concrete manner and
attracted some interest \cite{strominger,dietmar}. 

One of the interesting aspects of the conjectured dS/CFT correspondence is the
possibility of providing a quantum field theoretical description for
fields in de Sitter space. Such a goal is of obvious value in view of
the recent experimental observations for the existence of a positive
cosmological constant in our universe \cite{experimental}. Moreover,
if the dS/CFT 
correspondence resembles at all its successful cousin - the AdS/CFT
correspondence - then the possibility arises for a quantum
field theoretical 
description of the cosmological evolution
\cite{strominger2,minic}. Namely, considering that the
cosmological evolution in $D$ spacetime dimensions is modelled by a
(flat) metric of the FRW form 
\be
\label{frw0}
\rmd s^2 = -\rmd \tau^2 +{\cal R}^2(\tau)\rmd\bar{x}^2_{D-1}\,,
\ee
such that the Hubble constant 
\be
\label{hubble}
H = \frac{\dot{\cal R}}{\cal R} \rightarrow
H_{\pm}\,,\,\,\,\,\mbox{as}\,\,\,\,\, 
\tau\rightarrow \pm\infty\,,
\ee       
where the dot denotes differentiation with respect to $\tau$, then the
cosmological evolution could be viewed as a ``holographic'' RG 
flow between the two Euclidean CFTs dual to the asymptotic de Sitter
regimes at $\tau=\pm\infty$ of
(\ref{frw0}). Such an evolution from
the infinite past to the infinite future would  correspond to a
dual RG flow from the IR to the UV. 

Motivated by the above, as well as from the puzzling rareness of
string/M-theory compactifications to de Sitter space (see however
\cite{hull}), we consider
in this letter branes moving in 
an AdS$_{D+1}$ 
Schwarzschild black hole background. When the brane tension exceeds a
critical value we find solutions in which the induced
metric on the brane is of 
the general FRW form and approaches de Sitter space in the infinite
past and/or future. 

An interesting
class of solutions are those where the brane radius is always
greater than the event horizon of the AdS$_{D+1}$ black hole and stretches to
infinity both at the infinite past and future. In this case the brane
metric is asymptotically dS$_{D}$, with the same Hubble constant in
the past and future infinity. 
The
dS$_{D}$/CFT$_{D-1}$ correspondence applied to the asymptotic de Sitter
regimes of the 
above brane metrics implies the existence of dual CFTs whose 
central charges depend on the asymptotic Hubble constants. In our
case, the latter depend generically only on the brane 
tension, hence 
it seems that their values have nothing to do with any properties of the bulk
AdS$_{D+1}$ black hole. Phrased differently, the brane excitations induced
by the thermal bulk seem to be ``orthogonal'' to ``holographic'' de Sitter RG
flows. Nevertheless,  
these brane configurations break down when the mass of the bulk
AdS$_{D+1}$ black holes reaches a critical value that depends on the
brane tension. At this critical point, the solution interpolates
between a dS space and a space with Hubble constant zero and
corresponds to 
non-singular evolution on the brane (no Big Bang or Big Crunch).
It appears therefore that at this critical point the thermodynamics of the
AdS$_{D+1}$ black holes  
interferes with the
dS$_{D}$/CFT$_{D-1}$ correspondence. Indeed,
the central charge $c_{D-1}$ \cite{strominger2} of the CFT$_{D-1}$
is now related to the 
energy of the black hole, thus entering the
Cardy-Verlinde formula \cite{verlinde} that gives the entropy of the
thermal CFT$_D$ dual to the black hole 
\cite{witten}  as
\be
\label{result}
S = 2\pi \sqrt{\frac{\tilde{C}}{6}\left(
   A_D \,c_{D-1}
      -\frac{\tilde{C}}{ 24}\right)}\,.
\ee
The quantity $\tilde{C}$ is the generalized central charge of the thermal
      CFT$_D$ \cite{verlinde} and $A_D$ is a $D$-dependent
   constant. Since
      $\tilde{C}$ is the 
appropriate generalization to more than two dimensions of the number of
      degrees of freedom of the CFT$_D$ that are thermalized
   at the critical point 
      \cite{kps,kpsz}, it is tempting to relate $c_{D-1}$ to the
      maximum number of degrees of freedom on the brane that can be
      thermalized by a bulk black hole.

\section{Branes moving in AdS Schwarzschild black holes}

The equations describing the dynamics of codimension one surfaces
inside a bulk gravitational theory have been around for a long time
\cite{guth}. Due to the recent interest in brane world
scenaria, there exists by now a vast literature on the subject of
branes moving in bulk gravity. For our purposes, we take the simple
model of a brane moving in the background created by 
an AdS Schwarzschild black hole. This model has
been first studied in the context of AdS/CFT correspondence 
in \cite{verlinde2}. 
The $D+1$ gravitational action in the presence of a brane is
\be
\label{action}
S = \frac{1}{16\pi G_{D+1}}\int_{\cal M}(R -2\Lambda) +\frac{1}{8\pi
  G_{D+1}}\int_{\partial{\cal M}}\sqrt{\gamma}{\cal K} +\frac{\kappa}{8\pi
  G_{D+1}}\int_{\partial{\cal M}}\sqrt{\gamma}\,,
\ee
where ${\cal K}$ is the trace of the 
the extrinsic curvature ${\cal K}_{\m\n}=
  \gamma_{\m}^{\a}\gamma_{\n}^{\b} \nabla_{\a}\eta_{\b}$ taken with
  respect to the 
  induced metric on the brane $\gamma_{\m\n}=
  g_{\m\n}-\eta_{\m}\eta_{\n}$, $\kappa$ is related to
  the brane tension and $\eta_{\mu}$ is the unit normal vector to the
  brane.

A solution to the bulk equations of motion is given by the AdS
 Schwarzschild black hole in $D+1$ dimensions
\bea 
\label{adschwa}
\rmd s^2 &=& -h(r)\rmd t^2 +\frac{\rmd r^2}{h(r)} +r^2
\rmd\Omega_{D-1}^2 \,, \\
\label{hr} h(r) & = &
\frac{r^2}{L^2}+1-\frac{\omega_{D}M}{r^{D-2}}\,,\,\,\,\,\,
\omega_{D}=\frac{16\pi G_{D+1}}{(D-1)V_{D-1}}\,,
\eea
with $V_{D-1} =2\pi^{D/2}/\Gamma(D/2)$ and $M$ the mass of the black
hole.  In this normalization, the cosmological constant of 
$AdS_{D+1}$ is $\Lambda_{D+1}=-D(D-1)/2L^2$. To derive the equations
of motion we use Gaussian normal coordinates in the
vicinity of the brane
\be
\label{gaussian}
\rmd s^2 = \rmd\eta^2 + \gamma_{\m\n}\rmd x^{\m}\rmd x^{\n}\,,
\ee
with the position of the brane at $\eta=0$. We then
let the coordinates on the brane to be functions of the proper time
$\tau$ on
the brane as
$x^{\m}=(r(\tau),t(\tau),\bar{\theta})$, where the $\bar{\theta}$
denotes collectively the angular coordinates. 
The velocity of a static particle (or a piece of stress energy) on the
brane is 
\be
\label{q}
q^{\m} =\frac{\rmd x^{\m}}{\rmd \tau}= \left( \frac{\rmd t}{\rmd
    \tau}, \frac{\rmd r}{\rmd\tau}, \bar{0}\right) \,,\,\,\,\,\,
    q^{\mu}q_{\mu} =-1\,.
\ee
The second condition above is written as 
\be
\label{dtau}
\frac{1}{h(r)}\left(\frac{\rmd r}{\rmd \tau}\right)^2 -h(r)\left(\frac{\rmd
    t}{\rmd\tau}\right)^2 =-1\,,
\ee
which ensures that the induced metric $\gamma_{\m\n}$ on the brane is
    of the FRW type 
\be
\label{frw}
\rmd s^2_{brane} = -\rmd\tau^2 +r^2(\tau)\rmd\Omega_{D-1}^2\,.
\ee

In the coordinates (\ref{gaussian}) the normal vector to the brane
(i.e. at $\eta=0$),
pointing in the direction of increasing $r$, is simply
$\eta_{\m}=(1,0,\bar{0})$. To derive  the equations of motion,
however,  we need
the normal vector in the vicinity of the 
brane which is determined by
\be
\label{normalv}
\eta^{\m}q_{\m}=0\,,\,\,\,\,\,\eta^{\m}\eta_{\m}=1\,.
\ee
These are easily solved to give
\be
\label{eta}
\eta_{\mu} =
(-\dot r, \dot{t}, \bar{0})\,,
\,\,\,\,\dot{t}=\frac{\rmd   
  t}{\rmd \tau} \,,
\,\,\,\,\dot{r}=\frac{\rmd   
  r}{\rmd \tau}\,. 
\ee

Now, the equation of motion of the brane is given by the Israel
junction conditions \cite{guth} which, assuming that $\eta_{\m}$
points in the direction of increasing $r$, read
\be
\label{israel}
{\cal K}_{\m\n} = T_{\m\n}-T_{\rho}^{\rho}\frac{1}{D-1}\gamma_{\m\n}\,.
\ee
$T_{\m\n}$ is the energy momentum tensor on the brane which here
describes just its vacuum energy and is given by 
\be
\label{brane}
T_{\m\n} = -\kappa \gamma_{\m\n}\,.
\ee
Then (\ref{israel}) becomes
\be
\label{eom}
{\cal K}_{\m\n} = \frac{\kappa}{D-1}\gamma_{\m\n}\,.
\ee
To proceed we calculate the extrinsic curvature on the brane noticing
the relation 
\be
\label{gauss}
\partial_{\eta} = \eta^{\mu}\frac{\partial}{\partial
  x^{\mu}}\,.
\ee 
It suffices to consider just one of the angular components of ${\cal
  K}_{\m\n}$, e.g., the ${\cal K}_{\theta\theta}$ component in the
coordinate system (\ref{gaussian}) is
\be
{\cal K}_{\theta\theta} = -\Gamma_{\theta\theta}^{\eta} \eta_{\eta} =
\frac{1}{2} \eta^{\mu}\partial_{\mu}r^2 = rh(r)\dot{t}\,.
\ee
Then (\ref{eom}) yields
\be
\label{verleq}
\dot t = \frac{\kappa r}{(D-1)h(r)}\,.
\ee
The gravitational constant of the induced metric can be easily
calculated \cite{maeda} as
\be
\label{gravct}
\Lambda_{D} =
\frac{D-2}{D}\left(\Lambda_{D+1}+\frac{D}{2(D-1)}\kappa^2\right)
\equiv \frac{D-2}{2(D-1)}(\kappa^2-\kappa_c^2)\,,
\ee
where (\ref{gravct}) also defines the critical tension
$\kappa_c$. Finally, using (\ref{dtau}) we obtain from (\ref{verleq}) 
\be
\label{friedm}
H\equiv \left(\frac{\dot{r}}{r}\right)^2 =
\frac{2\Lambda_D}{(D-2)(D-1)}
-\frac{1}{r^2} +\frac{\omega_D M}{r^{D}}\,.
\ee
This equation describes the cosmological evolution for the brane
universe in the AdS Schwarzschild black hole background. Similar
equations have been studied in various braneworld related contexts 
in a number of references \cite{braneworlds}. 

\section{Asymptotically dS metrics on branes}

\subsection{The $D$=4 case}

We start our analysis of (\ref{friedm}) by considering in detail the
case  $D=4$ when we have
\be
\label{friedm4}
\dot r^2 = \frac{\Lambda_4}{3} r^2 - 1 + {\omega_4 M\over r^2}\,.
\ee
The solution when the brane cosmological constant is zero corresponds to
fine tuning the brane tension to $\kappa=\kappa_c$ and has been
studied in detail in \cite{verlinde}. For $0<\kappa<\kappa_c$ we
obtain negative cosmological 
constant on the brane.\footnote{We do not consider here the
  unclear case of having negative tension branes.}
In this case we denote $H_1^2 =-\Lambda_4/3 >0$ and notice that for
$M=0$, which corresponds to the bulk 
space being pure AdS, there exists no solution. For $M>0$, we obtain
in terms of $x=r^2$
\be
\label{ads}
\dot{x}^2 =-4H_1^2 (x-{\rm x}_-)(x-{\rm x}_+)\,,\,\,\,\,\, {\rm x}_{\pm} =
-\frac{1}{2H_1^2}\left[ 1\mp \sqrt{1+4\omega_4MH_1^2}\right]\,.
\ee  
Only ${\rm x}_+$ is positive, thus we have a bounded motion in the interval
$x(\tau) =r^2(\tau) \in (0,{\rm x}_+]$. Then, the solution is
\be
\label{adssol}
x(\tau)\equiv r^2(\tau) =
\frac{1}{2H_1^2}\left[\sqrt{1+4\omega_4MH_1^2}\cos(2H_1\tau)-1 \right] \,,
\ee
which describes an oscillating universe.

For $\kappa > \kappa_c$ we obtain a positive cosmological constant on
the brane. Denoting $H_2^2 =\Lambda_4/3$ we  
obtain from (\ref{friedm4}) 
\be\label{friedm2}
{1\over 4} \, \dot x^2 = H_2^2 x^2 - x + \omega_4 M = H_2^2
(x-x_-)(x-x_+)\,, 
\ee
where
\be 
\label{roots}
x_\pm = {1\over 2H_2^2} \left( 1\pm \sqrt{1-4\omega_4 M H_2^2} \right) \,.
\ee
We begin with the extremal case $M=0$, i.e., when the bulk is pure
AdS. The solution is then
\be 
r(\tau) = {1\over H_2} \; \cosh (H_2\tau) \,,
\ee
which in view of (\ref{frw}) describes a de Sitter universe. This case
has been studied from different perspectives, e.g., in
\cite{braneworlds,freedman}.  

If $4\omega_4 M H^2 < 1$, both roots (\ref{roots}) are positive. In
this case there 
are two possible orbits, one in the interval $(0,x_-]$ and another in
the interval $[x_+,\infty )$. Since one easily verifies that the black
hole horizon distance satisfies $r_H<x_-,x_+$, in the first orbit
above the
brane crosses the horizon at finite proper time, while in the second
orbit the brane never crosses the 
horizon. 
Eq.~(\ref{friedm2}) can be solved exactly.
After some algebra, we find that for $x\le x_-$,
\be r^2(\tau) = x(\tau) = {1\over 2H_2^2} \left( 1- \sqrt{1-4\omega_4 MH_2^2} \; \cosh
  (2H_2\tau) \right)\,,
\ee 
whereas for $x\ge x_+$, 
\be 
\label{orbit}
r^2(\tau) = x(\tau) = {1\over 2H_2^2} \left( 1+ \sqrt{1-4\omega_4
    MH_2^2} \; \cosh 
  (2H_2\tau) \right)\,.
\ee 
When $4\omega_4 M H_2^2 > 1$, we obtain the solution
\be 
\label{orbit3}
r^2(\tau) = x(\tau) = {1\over 2H_2^2} \left( 1+ \sqrt{4\omega_4
    MH_2^2-1} \; \sinh 
  (\pm 2H_2\tau) \right)\,.
\ee 
Finally, for $4\omega_4 M H_2^2 = 1$ we obtain the  solutions
\be 
r^2(\tau) = x(\tau) = {1\over 2H_2^2} \left( 1\pm e^{\pm 2H_2\tau}
\right)\,.
\ee
At this critical point there are two possible orbits one of which
interpolates between
a dS space of Hubble constant $H_2$ and a 
space of $H=0$. This corresponds to a
non-singular evolution (without a Big 
Bang or a Big Crunch, depending on the sign choice for $H_2$). The
various orbits are shown in Fig.~1.
\begin{figure}[ht]
\centering
\epsfysize=8cm
\begin{minipage}{16cm}
\centering
\psfrag{AAAAAA}{$4\omega_4 M H_2^2=1$}
\psfrag{BBBBBB}{$4\omega_4 M H_2^2<1$}
\psfrag{CCCCCC}{$4\omega_4 M H_2^2>1$}
\psfrag{DDDDDD}{$M=0$}
\psfrag{ttt}{$\tau$}
\psfrag{rrr}{$x$}
\psfrag{250}{$+\infty$}
\psfrag{-150}{$-\infty$}
\psfrag{-4}{$-\infty$}
\psfrag{4}{$+\infty$}

\includegraphics[width=16cm]{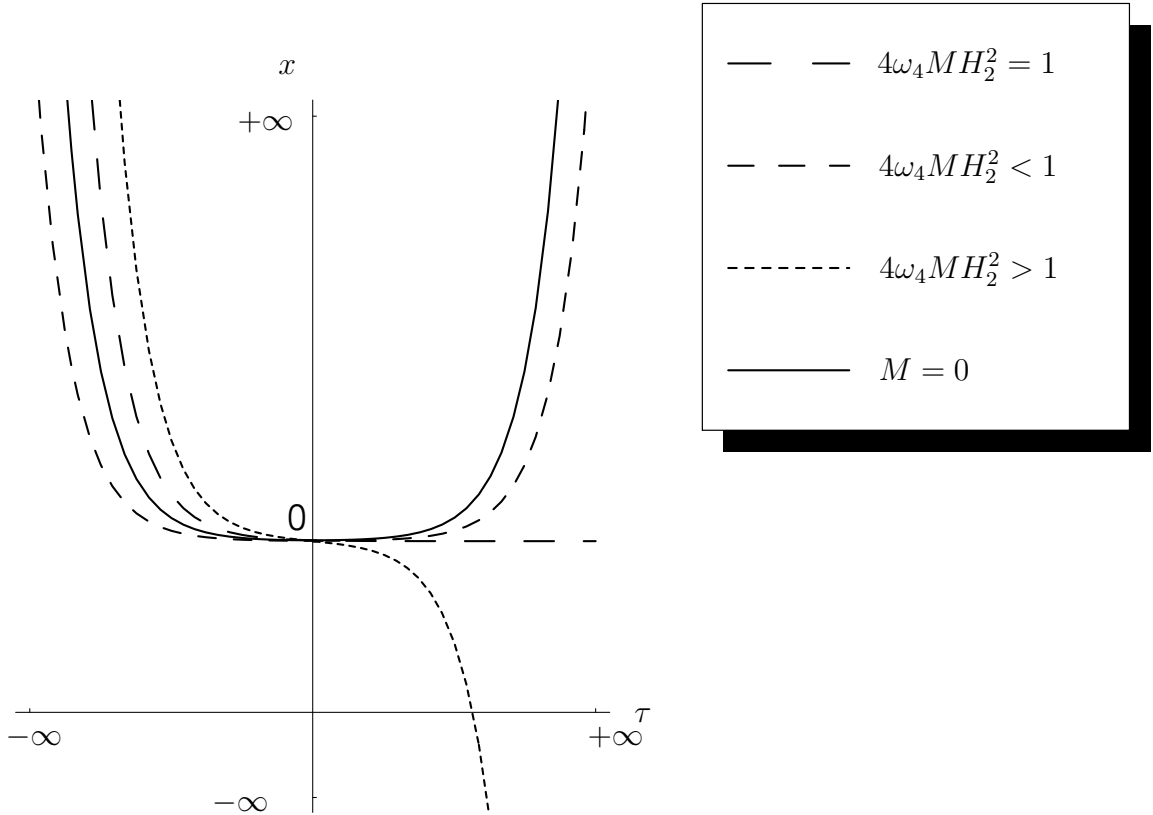}
\centering
\caption{The various brane orbits. For $4\omega_4 M H_2^2<1$ we depict
  only the $x>x_+$ orbit. For $4\omega_4 M
  H_2^2>1$ the brane ceases to exist  ($x<0$) at finite proper time.}
\end{minipage}
\end{figure}

From our perspective the most interesting solution is
(\ref{orbit}). In this case the brane 
metric (\ref{frw}) is asymptotically de Sitter both at the past and
future infinity and furthermore it never crosses the black hole
horizon. The brane 
observer sees a cosmological evolution during which the universe
starts very big at an inflationary de Sitter era, contracts to a minimum size
and then expands again to reach a de Sitter era in the infinite future.
Both the past and the future asymptotically de Sitter regimes have the
same Hubble constant $H_2$. In this sense, our brane universe is far
from describing the observed cosmological evolution where the initial
and final values of the Hubble constant differ by $\sim 10^{52}$ orders of
magnitude. Nevertheless, our solution provides a simple model to test
the recent ideas of dS$_4$/CFT$_3$ correspondence in the context of
the cosmological evolution \cite{strominger2,minic}. According to
such ideas, the asymptotically de Sitter regimes
in the infinite past and future are respectively dual to a
three-dimensional CFT on the sphere $S^3$ with radius $r$. The central
charge of this dual CFT$_3$ is given  
by \cite{strominger2}
\be
\label{ccharge3}
c_3 = \frac{a_{3}}{H_2^2 G_{4}}
\ee
where $G_4$ is the gravitational constant on the brane.  The
parameter $a_{3}$ is a 
constant.\footnote{The parameter $a_3$ can presumably be calculated
  from the correlators of the energy momentum tensor of the
  CFT$_3$. The latter can be calculated from graviton correlators in
  dS$_4$. Notice however, that 
the cosmological implementation above of the dS/CFT
      correspondence requires the definition of a central charge
      for an odd-dimensional CFT. Contrary to common misconceptions in
      the current literature, the notion of a quantity measuring the
      degrees of freedom coupled to a critical point is not
      necessarily connected to the conformal anomaly and exists in any
      dimension. For various 3-dimensional examples see
e.g. \cite{tassos}.} 
For general
$0\leq 4\omega_4 M H_2^2 <1$, we see from (\ref{gravct}) that $H_2$ is
independent of the  
black hole mass, hence the central
charge of the dual CFT$_3$ depends essentially only on the brane
tension $\kappa$. Therefore, although the presence of a black hole
in the bulk distorts the orbit and from purely de Sitter makes it
only asymptotically de Sitter, there does not seem to be any
interaction of the black hole degrees of freedom with the degrees of
freedom on the brane that are relevant to the boundary CFT$_3$,
i.e. this particular brane evolution is ``orthogonal'' to the RG flow
of the CFT$_3$. However, 
when the black hole mass reaches the critical value
\be M = M_c = {1\over 4\omega_4 H_2^2}\,,
\ee
then there seems to be an interference  between the degrees of
freedom of the black hole and those of the CFT$_3$. In this case,
starting, for example,  the evolution from a de Sitter regime in the
infinite past, the brane contracts and never re-expands again reaching
a spatially homogeneous space with zero Hubble constant in the infinite
future. From the point of view of the CFT$_3$ such an evolution
corresponds to a RG flow from an IR regime, where a finite number of
degrees of freedom are coupled, to an UV regime where the coupled
degrees of freedom are infinite. 

On the other hand, when $M>M_c$ we see from
(\ref{orbit3}) that $r^2(\tau)$ 
becomes negative at finite proper time which indicates that the brane
configuration ceases to exist. In that sense, $c_3$ in
(Eq.~(\ref{ccharge3})) provides 
an upper bound on the energy of black holes that can support brane
configurations stretching from the infinite past to the infinite
future. Indeed, using 
the equation $h(r_{H})=0$ that gives
the horizon distance $r_H$ of the black hole, the condition $M\le M_c$
translates to
\be
\label{critical}
c_3 =\frac{a_3}{H_2^2 G_4} \ge
\frac{a_3}{G_4}4r_{H}^2\left(1+\frac{r_H^2}{L^2}\right) \,.
\ee 
The equality in (\ref{critical}) associates a ``critical'' CFT$_3$ to an AdS$_5$ Schwarzschild
black hole of radius $L$ and horizon distance $r_H$. Apparently,  the
bulk AdS black hole does know something about  the ``critical''
theory whose central charge is given by (\ref{ccharge3}).
Using the standard brane-world relation
\be
\label{coupling}
G_{D+1} =\frac{L}{D-2}G_{D}\,,
\ee
we obtain from (\ref{critical})
\be
\label{critical1}
c_3 = a_3\frac{16}{3\pi}E_{crit}L \gtrsim a_3\frac{16}{3\pi} E L\,,
\ee
where $E$ is the energy of the black hole that supports a brane
cosmology and which is identified with the
total thermal energy of the dual CFT$_4$. $E_{crit}$ is the energy of
the black hole at the
critical point $M=M_c$. At this critical point, using the definition
of the Casimir entropy $S_C$
\be
\label{casimirE}
S_C = S\frac{L}{r_H}\equiv \frac{\pi}{6}\tilde{C}\,,
\ee
where $S$ is the entropy and $\tilde{C}$ the generalized central
charge \cite{verlinde,kps,kpsz}, we can write the entropy of the black
hole as
\be
\label{CV}
S =
2\pi\sqrt{\frac{\tilde{C}}{6}
  \left(\frac{\pi}{16 a_3}c_3-\frac{\tilde{C}}{24}\right)} \,.  
\ee 
Thus, the central charge of the CFT$_3$ enters the Cardy-Verlinde 
formula for the entropy of the thermal CFT$_4$ at the critical point $M=M_c$. 

\subsection{Generalization to $D>4$}

For $D>4$ we consider only the
case with a positive cosmological constant on the brane when
Eq.~(\ref{friedm4}) generalizes to
\be \dot r^2 = H_2^2r^2 - 1 + {\omega_D M\over r^{D-2}}\,, 
\ee
with $H_2^2 = {2\Lambda_D\over (D-2)(D-1)}$. 
This cannot be solved exactly, but we may demonstrate the qualitative features we encountered in
$D=4$. To this end, first rewrite the equation as
\be r^{D-2} \dot r^2 + f(r) =0\quad,\quad f(r) = -H_2^2r^D + r^{D-2} -
\omega_D M\,.
\ee
The potential $f(r)$ has a maximum at $r_{max}$, where $f'(r_{max}) = 0$, i.e.,
\be r_{max}^2 = {(D-2)\over DH_2^2} \,.
\ee 
As in $D=4$, we have three cases:

If $f(r_{max}) > 0$, then there are two orbits, one of which approaches $r\to \infty$ in the infinite
past and future. We obtain $\dot r^2 \approx H_2^2 r^2$, therefore,
\be r \sim e^{\pm H_2\tau} \,,
\ee
and the brane interpolates between dS spaces of Hubble constant $H_2$.

If $f(r_{max}) < 0$, then the brane starts from a dS space of Hubble constant $H$ in the
infinite past and falls into the black hole singularity.

At the critical point $f(r_{max}) = 0$, the mass of the black hole becomes
$M=M_c$, where $M_c$ satisfies
\be\label{eqHM}
{1\over H_2^{D-2}} = {D-2\over 2} \left( {D\over D-2} \right)^{\frac{D}{2}}
\omega_D M_c \,.
\ee 
The brane interpolates between a dS space of Hubble constant $H_2$ in the infinite future and a finite space of
radius $r=r_{max}$ in the infinite past. There is also another orbit that approaches the latter
point starting from the singularity $r=0$. To see the behaviour near
$r\to r_{max}$, expand 
\be f(r) = \half f'' (r_{max}) (r-r_{max})^2 + \dots = (D-2)
r_{max}^{D-4} (r-r_{max})^2 + \dots\,,\ee 
Therefore, 
\be \dot r^2 = - {f(r)\over r^{D-2}} \approx - {(D-2)\over
  r_{max}^2}\, (r-r_{max})^2 
= - DH^2\, (r-r_{max})^2\,,
\ee
whose solution is
\be
r -r_{max} \sim e^{\pm \sqrt D\, H}\,.
\ee
In this case, using (\ref{eqHM}) and the equation for the black hole
horizon, the central charge of the CFT$_{D-1}$ dual to the
asymptotic dS$_{D}$ regime is
\be 
\label{cchargeD}
c_{D-1} = \frac{a_{D-1}}{H^{D-2}G_{D}} =
\frac{D-2}{2}\left(\frac{D}{D-2}\right)^{\frac{D}{2}} r_{H}^{D-2}\left(1+
  \frac{r_H^2}{L^2}\right) \frac{a_{D-1}}{G_D}\,,
\ee
which using (\ref{coupling}) and the formula that gives the energy of
the bulk black hole \cite{witten} can be written as
\be
\label{cchargeD1}
c_{D-1}  \frac{8\pi a_{D-1}}{V_{D-1}(D-1)}\left(
  \frac{D}{D-2}\right)^{\frac{D}{2}} E_{crit}L\,.
\ee
The $D$-dependent parameter $a_{D-1}$ is a constant (see Footnote 1).
Away from the critical point, the condition for the existence of asymptotic
dS spaces is $E<E_{crit}$. Thus $c_{D-1}$ provides an upper bound for the energy of the
black hole which generalizes (\ref{critical1}).
Moreover, the Cardy-Verlinde formula
  for the entropy of the thermal CFT$_D$ is written  at the critical
  point $M=M_c$ as
\be
\label{resultagain}
S = 2\pi \sqrt{\frac{\tilde{C}}{6}\left(
      \frac{V_{D-1}}{8\pi}\left(\frac{D-2}{D}
      \right)^{\frac{D}{2}}\!\! \frac{c_{D-1}}{a_{D-1}}
      -\frac{\tilde{C}}{ 24}\right)}
\ee
generalising~(\ref{CV}).

\section{Discussion of the results}

The recent experimental observations pointing towards the existence of
a positive cosmological constant in our universe has brought up the
question of studying  
quantum field theory in de Sitter space. In that direction, the
conjectured dS/CFT  
correspondence is an interesting development. A
further  possibility arising in this context is the description of the
cosmological evolution via a RG flow between 
Euclidean CFTs dual to the asymptotic dS regimes of the infinite past
and future. With the above in mind, it seems natural to look
for realizations of the dS/CFT correspondence in spaces embedded into
higher dimensional bulk manifolds. In this way one might hope to use the
full range of the existing string theory and supergravity solutions for
the bulk to study the dS/CFT correspondence realized on branes.

In this letter, we initiate such an approach to dS/CFT correspondence
by studying the simple model of a  brane moving in the vicinity of an AdS
Schwarzschild black hole. When 
the tension is greater than a certain critical value, the
metric on the brane is of a generic FRW form which is asymptotically
de Sitter in the infinite past and/or future. We argue that such a
simplified configuration is a good starting point to study properties
of the dS/CFT correspondence. An interesting question is
whether the bulk theory is at all related to the properties of the CFT(s) dual
to the asymptotic dS regimes. Such a possibility would then allow one
to use information from the well established AdS/CFT correspondence
to study the dS/CFT correspondence. 

Our analysis shows that although  a black hole in the bulk distorts
the orbit of a brane moving into it, in general it does not
seem to interfere with the CFTs dual to the asymptotic dS regimes on the
brane. However, for a critical value of the black hole mass
the asymptotic properties of the brane metric are dramatically altered
and the brane interpolates between a dS space in the infinite future/past and a
space of zero Hubble constant in the infinite past/future. This is a
non-singular 
evolution (without a Big Bang/Big Crunch). Such a critical point defines
an upper bound on the energy of a Schwarzschild AdS black hole that
can support an 
asymptotically de Sitter brane cosmology. 
We view this as an indication that the black hole knows something
about the CFT dual to the dS regimes. This is further supported by
observing that at the critical point the central charge of the
CFT$_{D-1}$ enters 
the Cardy-Verlinde entropy formula (\ref{result}) of the thermal
CFT$_D$ dual to the 
black hole.  Such an observation indicates that the central
charge of the CFT$_{D-1}$ may be related to a quantity of the thermal
CFT$_D$. From the Cardy-Verlinde formula (\ref{result}) one is tempted
to relate $c_{D-1}$ to the maximum number of degrees of freedom 
on the brane that can be thermalized by the bulk black hole, for
temperatures above the Hawking-Page transition point \cite{witten}. 

Our arguments above require the existence of a well-defined central charge
for a $D$-dimensional CFT. In the context of the dS/CFT
correspondence, this means a general formula for the constant $a_{D-1}$
in (\ref{cchargeD}). Such a formula is, of course, not known for
$D>2$. In even dimensions the central charges are usually related to
conformal anomalies and this makes their study relatively simple.
Nevertheless, central charges have been defined in odd
dimensions \cite{tassos} and could be used to study properties of the
dS/CFT correspondence in the cosmological evolution setting.

The simple model presented here shows the possibility of studying
properties of the dS/CFT correspondence in embedded spaces
(i.e. branes), using results of the AdS/CFT correspondence. One could
extend our result to asymptotically flat brane cosmologies in which
case we expect that one should use the Cardy-Verlinde entropy formula
for black holes with flat horizons discussed in \cite{kpsz}. Notice
that at the critical point the evolution on the brane leads to an
infinite ratio of the  
Hubble constants in the infinite future and infinite past, respectively. It
would be interesting to perturb this symmetric state and obtain a large but
finite ratio of Hubble constants as observed in our universe. In that
sense, a natural
extension of our studies would be to consider more complicated black
hole solutions of supergravity and study the influence of the bulk
fields on possible asymptotic dS regimes on the brane. Introducing
explicitly thermal conformal fields on the brane would also be a
possible way forward. In any case, the idea of viewing the cosmological
expansion as a RG flow is by itself interesting and worths further
study.

\subsection*{Acknowledgements}
A. C. Petkou would like to acknowledge discussions  with
Y. Oz and J. Pawelczyk.

\end{document}